# An Image Registration Based Technique for Noninvasive Vascular Elastography

Sina Valizadeh[a,b], Bahador Makkiabadi[a,b], Alireza Mirbagheri[a,b], Mehdi Soozande[a,b], Rayyan Manwar[d], Moein Mozaffarzadeh[b,c], and Mohammadreza Nasiriavanaki[d]

[a]Department of Medical Physics and Biomedical Engineering, School of Medicine, Tehran University of Medical Sciences, Tehran, Iran.
[b]Research Center for Biomedical Technologies and Robotics (RCBTR), Institute for Advanced Medical Technologies (IAMT), Tehran, Iran.
[c]Department of Biomedical Engineering, Tarbiat Modares University, Tehran, Iran.
[d]Wayne State University, Bioengineering Department, Detroit, Michigan, United States.

## ABSTRACT

Non-invasive vascular elastography is an emerging technique in vascular tissue imaging. During the past decades, several techniques have been suggested to estimate the tissue elasticity by measuring the displacement of the Carotid vessel wall. Cross correlation-based methods are the most prevalent approaches to measure the strain exerted in the wall vessel by the blood pressure. In the case of a low pressure, the displacement is too small to be apparent in ultrasound imaging, especially in the regions far from the center of the vessel, causing a high error of displacement measurement. On the other hand, increasing the compression leads to a relatively large displacement in the regions near the center, which reduces the performance of the cross correlation-based methods. In this study, a non-rigid image registration-based technique is proposed to measure the tissue displacement for a relatively large compression. The results show that the error of the displacement measurement obtained by the proposed method is reduced by increasing the amount of compression while the error of the cross correlation-based method rises for a relatively large compression. We also used the synthetic aperture imaging method, benefiting the directivity diagram, to improve the image quality, especially in the superficial regions. The best relative root-mean-square error (RMSE) of the proposed method and the adaptive cross correlation method were 4.5% and 6%, respectively. Consequently, the proposed algorithm outperforms the conventional method and reduces the relative RMSE by 25%.

## 1. INTRODUCTION

Pathological conditions often induce changes in the biological tissue stiffness. It necessitates using the elasticity imaging, which aims to outline the elastic properties of the soft tissues.[1] Recently, several groups have proposed some approaches for non-invasively characterization of the superficial arteries. Standard two-dimensional (2D) correlation technique is usually used to compute the displacement map in the consecutive images.[2] 2D cross correlation method is a well-known algorithm to measure the tissue displacement. In the case of a low stress, the displacement of speckles which are at the long distance from the center of the vessel is too small to be detected by ultrasound imaging. In the case of a large compression, the search region would be large, and therefore, the cross-correlation method results in an erroneous displacement estimation due to a phenomenon called peak hopping. To reduce the error of the displacement estimation, an adaptive 2D search method was developed by Korukonda in 2011.[3] This algorithm improves the displacement estimation for a relatively large compression, but still does not have enough precision due to the deformation of the search window. Hence, non-rigid image registration based technique which evolves the deformation might yield a more reliable performance. Also, researchers have measured the axial and lateral components of the displacement with 2D echo tracking and optical-flow-based

---

displacement estimation techniques.[4,5] All the proposed techniques have produced a reliable axial displacement estimation, but the quality of the lateral displacement estimation is limited due to the low lateral resolution of the imaging systems. In 2012, Korukonda suggested a synthetic aperture (SA) based ultrasound imaging method to improve the lateral resolution.[6] The poor image quality in the regions near the transducer is a significant drawback of the SA technique. This limitation has been neglected in the previous works. In this work, a non-rigid image registration technique is suggested to measure the tissue displacement in a 2D plane. The strain map can be measured with an acceptable precision in the near and far distances from the center of the vessel by applying a relatively large stress. The SA method is used to acquire the ultrasound images with a high axial and lateral resolution. Since the most of the vessels are superficial, it is vital for the imaging system to have an enough resolution in the superficial region. Here, we used the directivity diagram to improve the lateral resolution in the regions near the transducers.

## 2. MATERIALS AND METHODS

For the simulations, a 3D model of vessel has been constructed in Solidworks (3D cad modelling software) where the inner radius, the outer radius and the length are 1.5 $mm$, 6 $mm$ and 100 $mm$, respectively. The cad model has been imported to Ansys (a finite element method (FEM) software) where the module of elasticity, the poison ration and the pressure difference are 40 $kPa$, 0.495 and 700 $Pa$, respectively. The SA method is implemented for ultrasound imaging.[7,8] We used the delay-and-sum (DAS) technique to reconstruct the SA images.[9,10] Using an N-element array, for each point in an image, the A-scan signals can be expressed as (1):

$$A(r,\theta) = \sum_{m=1}^{N-1}\sum_{n=1}^{N-1} y_{m,n}(\frac{2r}{c} - \tau_{m,n}), \tag{1}$$

where $y_{m,n}(t)$ is the echo signal and $\tau_{m,n}$ is the beamforming delay for the $(m,n)$ receive and transmit element combination.[11,12] Here, a modified SA imaging algorithm which uses the directivity diagram is used.[13] The modified version improves the lateral resolution and the image quality in the regions near the transducer. The modified SA can be written as follows:

$$A(r,\theta) = \sum_{m=1}^{N-1}\sum_{n=1}^{N-1} f(\theta_m)f(\theta_n) y_{m,n}(\frac{2r}{c} - \tau_{m,n}), \tag{2}$$

where $\theta$ is the corresponding observation angle for each transmit-receive pair. $f(\theta)$ can be calculated by (3):

$$f(\theta) = \frac{sin(\pi d/\lambda sin(\theta))}{\pi d/\lambda sin(\theta)} cos(\theta), \tag{3}$$

where $d$ and $\lambda$ are the width of element and the wavelength, respectively. A non-rigid image registration based method is applied to ultrasound images to measure the displacement of each particles in the vessel wall. Medical image registration toolbox i.e. MIRT, is used to implement the image registration algorithm.[14–19] MIRT meshes the reference and the target images first. Then, using a similarity criterion, the displacement of the grid points in the target image to the reference image is estimated. An adaptive cross correlation algorithm is also implemented for comparison.[3,20] The relative root of mean square error (RMSE) is used to evaluate the displacement estimation techniques (4).

$$RMSE(\%) = \sqrt{mean\Big(\big(\frac{Dis_{FEM} - Dis_{MIRT}}{Dis_{FEM}}\big)^2\Big)} * 100, \tag{4}$$

where, $Dis_{MIRT}$ is the amount of estimated displacement in each pixel of the ultrasound image, $Dis_{FEM}$ is the amount of calculated displacement by FEM corresponding to ultrasound image pixels, and $n$ is the number of pixels located at the region of the vessel wall.

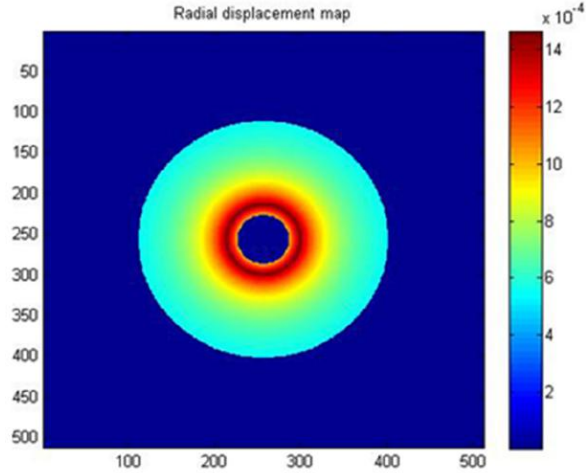
Figure 1: The map of the displacement obtained by FEM.

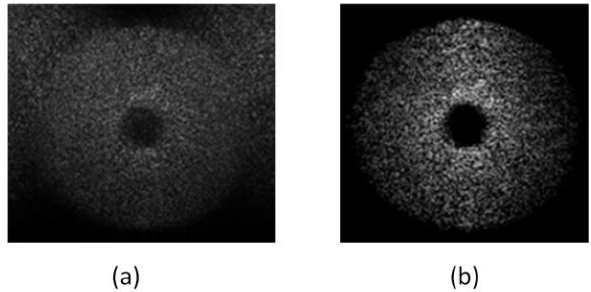

Figure 2: The ultrasound images of the simulated vessel using (a) Conventional SA, (b) Modified SA.

## 3. RESULTS

The map of the displacement obtained by FEM is illustrated in Figure 1. We used the acoustic field simulation program (Field II), to simulate the SA ultrasound imaging.[21,22] The number of elements, central frequency, element height, element width, element pitch, kerf and sampling rate are 128, 7.5 $MHz$, 4.0 $mm$, 0.2789 $mm$, 0.3048 $mm$, 0.025 $mm$ and 100 $MHz$, respectively. The center of the acoustic phantom of the vessel is placed in a distance of 1 $cm$ from the transducer surface in order to imitate a superficial vessel. The resulted images of the vessel acquired by the SA method without/with applying the directivity diagram are shown in Figure 2(a) and Figure 2(b), respectively. The parameters of MIRT are detailed in Table 1.The resulting map of the displacement estimated by MIRT and adaptive cross correlation technique are illustrated in Figure 3 and Figure 4, respectively. The performance of the proposed algorithm and the adaptive cross correlation method are evaluated using the RMSE (see Figure 5). Several techniques are proposed to measure the displacement and consequently the strain in NIVE. Each has some advantages and disadvantages. Adaptive cross correlation which is exerted in NIVE, by Korukonda, figured out some restrictions of the former methods.[6] However, the most significant drawback of this method is its low performance for the large displacement estimation which confine this method to just a small compression in which the displacement is too small in the region far from the center of vessel. So, the imaging system cannot depict it with enough precision. Therefore, a large error occurs in these regions, leading to a low performance of this algorithm. Consequently, the lower bound of the strain is restricted by the resolution of the imaging system while the upper bound is determined by the displacement estimation algorithm.

## 4. DISCUSSION AND CONCLUSION

In this study, a non-rigid image registration based technique is used to measure the displacement in the case of relatively large strain. The performance of proposed method and the adaptive cross correlation technique would be improved when the applied pressure is increased because the displacement in the far regions from the center of vessel becomes apparent and can be measured (see Figure 3 and Figure 4). However, for a relatively high pressure which induces the strain of 3% or more, the adaptive cross correlation is failed in the estimation of a large displacement in the near regions. Thus, its performance begins to reduce for a large compression while the performance of the proposed method is still improving (see Figure 5). In addition to the image processing approach which has a significant influence on the precision of the displacement estimation, the spatial resolution of the imaging system is also a determinative parameter. To overcome the poor lateral resolution of the conventional ultrasound imaging which affect the precision of the displacement measurement in lateral direction, SA based methods are suggested. Whereas the most vessels are superficial, the imaging systems are required to have a proper resolution in this area. The main disadvantage of the SA method is its poor lateral resolution in the region near the transducer. This restriction is has not been taken into consideration in the previous works and vessels are assumed to be in an enough distance from the transducer. In this work the directivity diagram is applied to the SA ultrasound imaging to improve the lateral resolution in the superficial regions (see Figure 2). In conclusion, a non-rigid image registration based technique is used to measure the displacement in NIVE. The simulation results show that the proposed technique outperforms the adaptive cross correlation algorithm and reduces the relative RMSE for about 25%. The best relative RMSE of the proposed method is 4.5% in the strain of 10% while the best relative RMSE of the adaptive cross correlation method is 6% in the strain of 2% and rises to 17.4% in the strain of 10% (see Figure 5). So, the introduced algorithm seems promising and encourages for further practical considerations.

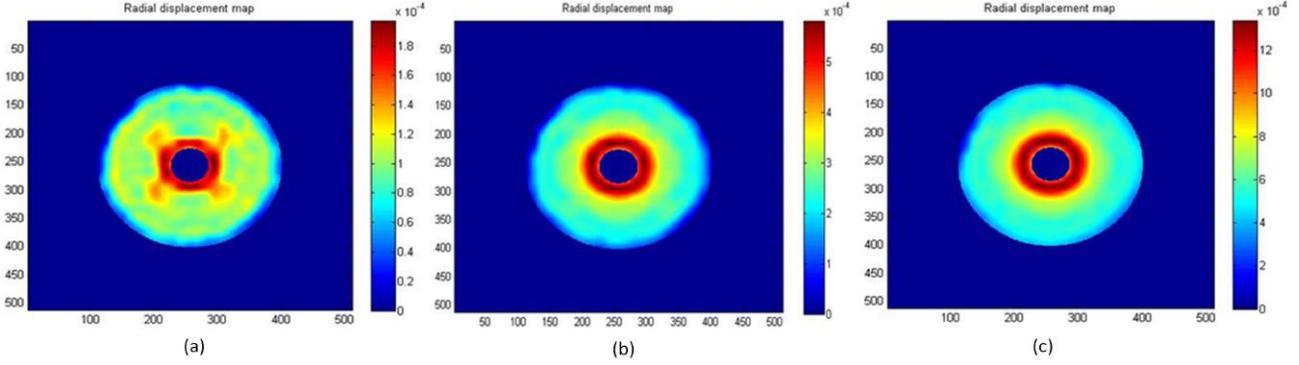

Figure 3: The map of displacement estimated by MIRT for different strain. (a) Strain of 0.1 %, (b) strain of 1%, and (c) strain of 10%.

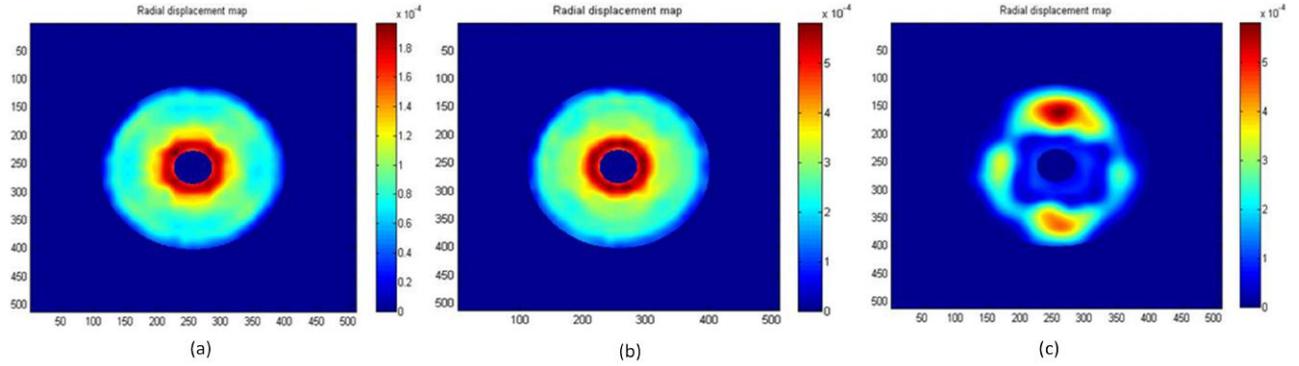

Figure 4: The map of displacement estimated by adaptive cross correlation for different strain. (a) Strain of 0.1%, (b) strain of 1%, and (c) strain of 10%.

## ACKNOWLEDGMENT

This study was part of a M.S. thesis supported by Tehran University of Medical Sciences; Grant NO. 93-02-30-2613.

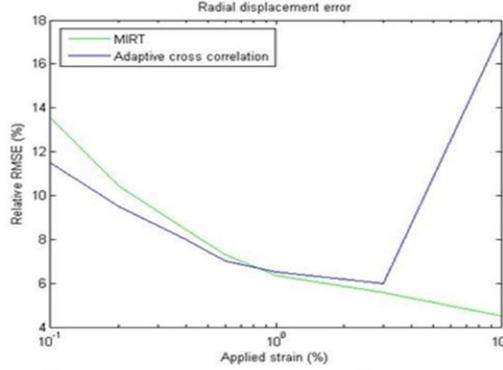

Figure 5: The relative RMSE of MIRT and adaptive cross.

Table 1: The parameters of MIRT for the displacement estimation.

| Parameter | Value |
|---|---|
| Similarity measure | SSD |
| Hierarchical level | 3 |
| Mesh window size | 15 |
| Transformation regularization weight | 0.05 |
| Maximum number of iteration | 50 |
| Tolerance (Stopping criterion) | $10^{-52}$ |
| Initial optimization step size | 1 |
| Annealing rate | 0.8 |